\documentclass[twocolumn]{aastex62}

\def\kms{\ifmmode{\rm km\thinspace s^{-1}}\else km\thinspace s$^{-1}$\fi}
\def\ms{\ifmmode{\rm m\thinspace s^{-1}}\else m\thinspace s$^{-1}$\fi}

\graphicspath{{./}{figures/}}

\received{\today}
\revised{xx, 2018}
\accepted{xx, 2018}
\submitjournal{AAS Journals}

\shorttitle{M$_{V}$ vs. Inclination in NL}
\shortauthors{Howell \& Mason}

\begin{document}

\title{On the M$_{V}$ -- Inclination Relationship for Nova-like Variables}
\correspondingauthor{Steve B. Howell}
\email{steve.b.howell@nasa.gov}

\author{Steve~B.~Howell}
\affil{NASA Ames Research Center, Moffett Field, CA 94035, USA}

\author{Elena~Mason}
\affiliation{INAF-OATS, Via G.B. Tiepolo 11, 34143, Trieste, IT}



\begin{abstract} 
Using a sample of Nova-like stars from the Ritter and Kolb catalog, we examine the relationship between their Gaia determined absolute magnitude and the inclination of the binary system. 
Webbink et al. (1987) derived a relationship between these two variables that provides a good fit and allows differentiation between $\dot{M}$ (and possibly M$_{WD}$) as a function of inclination. We show that the spread in M$_V$, at a given $i$, is dominated by the mass transfer rate with only a small dependence on the white dwarf mass. The validated relation shows that present-day theoretical population studies of cataclysmic variables as well as model fits to observational data yield mass transfer rates and white dwarf masses consistent with the Gaia derived M$_V$ for the nova-like stars.
\end{abstract}



\section{Introduction} \label{sec:intro}
There have been a number of previous works that attempted to discern the relationship between the absolute magnitude (M$_V$) of an accretion disk bearing binary star (i.e., a cataclysmic variable) and its system inclination. In principle, such a relationship is easy to imagine since seeing the accretion disk face-on will make the binary system substantially brighter than seeing only the thin edge of cooler material. From both first principles (Paczynski \& Schwarzenberg-Czerny, 1980; Mayo et al., 1980) and observational results mixed with theory (Warner, 1986; Webbink et al., 1987), astronomers have produced a number of empirical relationships  for M$_V$ vs. $i$ as an aid in the quest to understand the disk structure and distances to such binaries.  More recently, general values and equations, based on the above accepted paradigms pervade the literature and are casually used and excepted as correct for individual or classes of stars (e.g., Warner 1995, Patterson 2011; Ramsay 2017). 

The original relationships mentioned above were developed for specific cases; U Gem, UX UMa-like disks, recurrent novae, and nova remnants.  
Most of these types of cataclysmic variable generally have one thing in common -  their accretion disk is believed to be (or was modeled as) optically thick with the disk dominating the light output in the visible part of the spectrum. However, these past studies used a mixture of system types, orbital periods, and techniques to formulate their relationships.

In this paper, we revisit the connection between a bright disk system's absolute magnitude and its inclination. We use a model-independent approach based on a set of nova-like stars and new results from Gaia data release two (DR2). We assume that our stars M$_V$ is dominated by the light from the optically thick disk and that, for a constrained range in orbital period, their system properties are similar such that the disk inclination will dominate M$_V$;
white dwarf mass, $q$, and mass transfer rate being second order effects. There is, of course, much observational evidence that disks of nova-like systems are optically thick (e.g. Bisol et al. 2012, Baptista et al. 1995a,b). Our assumptions are well founded based on theoretical results (e.g., Howell et al., 2001; Kolb \& Baraffe 2000) and the literature reviewed later on.   

We discuss our sample in the next section, review some of the original relationships and highlight their features in \S3. Finally we summarize our results, and as a spoiler, we find that the old, venerable relationships work pretty well.

\section{The Nova-like Sample}
\begin{deluxetable*}{lrccccccc}
\tablecaption{The sample of NL with known orbital inclination extracted from the Ritter \& Kolb catalog and cross correlated with the Gaia DR2 and the Gaia TAP service of the Astronomisches Rechen Institut (ARI, Bailer-Jones et al. 2018).  \label{tab:table}
}
\tablehead{
\colhead{name} & \colhead{Gaia ID} & \colhead{d} & \colhead{d$_{min/max}$} & \colhead{M$_V$} & \colhead{V} & \colhead{$i\pm\sigma_i$} & \colhead{P$_{orb}$} & $i$ ref \\ 
 &  & (pc) & (pc) & (mag) & (mag) & ($^o$) & (hr) & \\ 
}
\startdata
    WX Ari &   22272497806324480 &  664 &  605/735 &  6.2 &  15.3  &  72-82$^{a}$ &     3.344 & Rodriguez-Gill+2000  \\
    RW Tri &  130692247044752784 &  312 &  308/317 &  5.0 &  12.5  &  70.5$\pm$2.5 &    5.565 & Smak 1995\\
J0107+4845 &  401879681868136704 &  748 &  714/786 &  5.5 &  14.9  &  81.4$\pm$0.1 &    4.646 & Khruzina+2013 \\
    DW UMa &  855119196836523008 &  577 &  565/590 &  4.8 &  13.6  &  82$\pm$4 &     3.279 & Araujo-Betancor+2003 \\
J0809+3814 &  908714959852556672 & 1222 & 1153/1300 &  5.2 &  15.6  &  65$\pm$5 &     3.217 & Linnell+2007b \\ 
  V482 Cam & 1108037726271701120 &  547 &  535/560 &  7.0 &  15.7  &  85$\pm$4 &     3.207 &  Rodriguez-Gil+2004 \\
    LX Ser & 1209876314302933504 &  486 &  476/496 &  5.7 &  14.1  &  80$\pm$3 &     3.802 & Magnuson 1984 \\
    UX UMa & 1559987685901122560 &  295 &  293/297 &  5.2 &  12.5  &  73$\pm$1.8 &      4.72 & Smak 1994 \\
    CM Del & 1815021160316471808 &  403 &  398/408 &  5.4 &  13.4  &  73$\pm$47&     3.888 &  R\&K$\dagger$  \\ 
  V425 Cas & 1996248233085177600 &  886 &  862/911 &  4.8 &  14.5  &  25$\pm$9 &      3.59 &  R\&K$\dagger$  \\
 V1776 Cyg & 2083145484587589632 & 1057 &  993/1130 &  6.1 &  16.2  &  75$^{+2}_{-1}$ &     3.954 & Garanvich+1990 \\
    MV Lyr & 2106069275529926400 &  493 &  481/505 &  3.3 &  11.8  &  7$\pm$1 &     3.176 & Linnell+2005  \\
    VY Scl & 2329317895999827968 &  630 &  607/654 &  3.1 &  12.1  &  15$\pm$10 &     5.575 & Schmidtobreick+2018\\
    VZ Scl & 2337436792938619392 &  552 &  534/571 &  6.9 &  15.6  &  76-90$^{a}$ &     3.471 &  Odonogue+1987\\
 0220+0603 & 2517357336654841856 &  717 &  656/790 &  7.0 &  16.3  &  79.1-79.7 &     3.581 &  Rodriguez-Gil+2015 \\
    UU Aqr & 2675351827511262720 &  254 &  249/259 &  6.3 &  13.3  &  78$\pm$2 &     3.931 & Baptista+1996   \\
    KR Aur & 3436435910858051072 &  451 &  377/563 &  4.4 &  12.7  &  20-40$^{a}$ &     3.907 &  Hutching+1983 \\
    RW Sex & 3769067109159365120 &  235 &  230/240 &  3.0 &   9.9  &  34$\pm$2 &     5.882 &  Vande Putte+2003\\
 V1315 Aql & 4313192491505026560 &  443 &  437/449 &  6.1 &  14.3  &  82.1$\pm$3.6 &     3.353 &  Dhillon+1991 \\
  V380 Oph & 4474002634076551680 &  667 &  635/702 &  5.2 &  14.3  &  42$\pm$13 &     3.699 & R\&K$\dagger$ \\
    RR Pic & 5477422099543151616 &  504 &  496/512 &  3.5 &  12.0  &  60-80$^{a}$ &     3.481 &  Riberio+2006  \\
    IX Vel & 5515820034889609216 &   90 &   90/91  &  4.2 &   9.0  &  57$\pm$2 &     4.654 & Linnell+2007a \\
  V347 Pup & 5553468275089335296 &  293 &  292/295 &  6.1 &  13.4  &  84.0$\pm$2.3 &     5.566 &  Thoroughgood+2005  \\
    OY Ara & 5931112341266391040 & 3175 & 2064/5573 &  6.2 &  18.7  &  74.4$\pm$1.3 &    3.731 & Zao+1997   \\ 
\enddata

$\dagger$ Inclination value (and error) are from the Ritter and Kolb catalogue referring to the unpublished Shafter (1983) PhD thesis. 
$^{a}$ These stars only have inclination ranges. 
\end{deluxetable*}

We selected our sample of nova-like stars from the 2014 version of the Kolb \& Ritter CV Catalogue\footnote{
https://heasarc.gsfc.nasa.gov/W3Browse/all/rittercv.html}. {\bf We selected all stars listed as `NL` within the orbital period range of 3-6 hr. This search gave us 87 stars, of which 25 had inclination and other pertinent system information that made up our sample list. While we are aware that this sample is not complete, it is sufficient to demonstrate the goals of this paper. 
Since the inclination, $i$, is the parameter of interest here, we have checked its value for each of the objects in the sample. In particular we made sure that each inclination value that we adopted (either as reported in the Ritter and Kolb catalog or by some other source) was derived by careful light curve modeling (through multiple sources) and/or spectroscopic/radial velocity studies. In the few cases where the error was not listed in the published work, we estimated it ourselves using the published information (e.g. either stellar component mass uncertainties or the presented graphic solution). Four objects have only a range constraint for their inclination.}

For these 25 stars, we queried the Gaia DR2 in the Heidelberg ARI's Gaia mirror site (http://gaia.ari.uni-heidelberg.de/) first to get the GDR2 IDs number of each source and then the corresponding best distance as estimated by Bailer-Jones et al. (2018).The distance information was derived by the Gaia team (Gaia Collaboration et al. 2018; Luri et al. 2018) and includes their error estimates (min/max distance), manifested in our M$_V$ values as uncertainties. {\bf The Gaia team did extensive tests of their distance determinations throughout the sky in terms of the correctness and uncertainties (See Luri et al. 2018; Clementini et al., 2018). No ARI distance was available for SW Sex, thus it was removed from our sample. Table 1 lists our final sample of 24 stars.} 
OY Ara shows a larger than usual Gaia distance uncertainty due to the object's faint apparent magnitude. 

Using the catalog high state $V$ magnitude for each NL, we calculated its M$_V$. We note that most of our objects have E(B-V) consistent with zero (see the literature discussed below) thus we ignore reddening in the distance determinations. 
Table 1 lists the distance information calculated by the Gaia team, including most likely and limiting values, and our calculated absolute magnitude values. We find no correlation of M$_V$ with orbital period or distance as expected, but given the generally similar distances (200-800 pc) of the stars in the sample, the brighter absolute magnitude stars are also generally brighter in V magnitude as well. 

According to theory (e.g., Kolb \& Baraffe, 2000; Howell et al., 2001; Kalomeni et al., 2016), these stars should have mass transfer rates between 10$^{-8}$ and 10$^{-9}$ M$_\odot$/yr, providing them with a bright, steady optically thick accretion disk yielding $\sim$100\% of the optical light from the system.

\section{Absolute Magnitude vs. Inclination}
\begin{figure}
\label{f1}
\includegraphics[angle=0,scale=0.42,keepaspectratio=true]{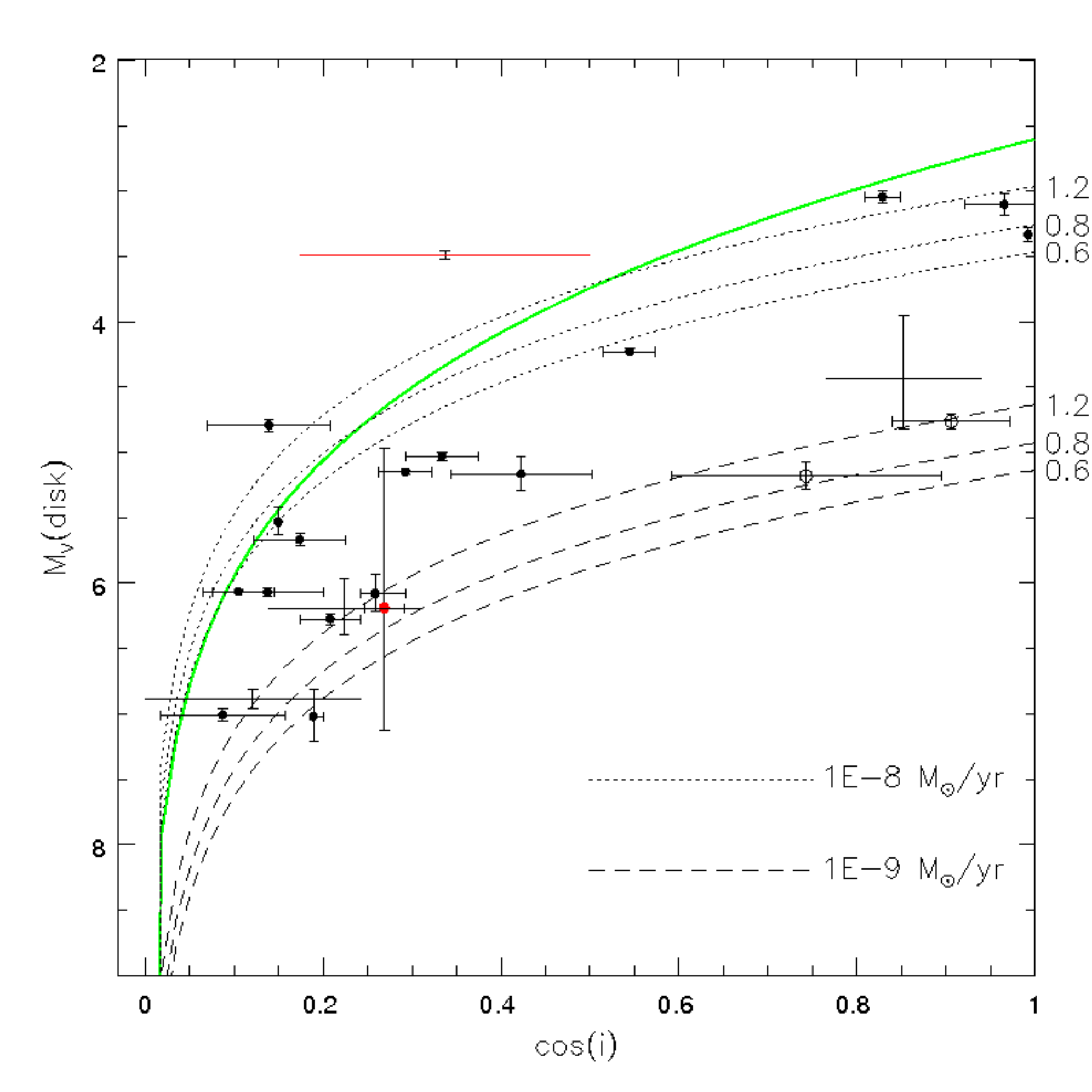}
\caption{M$_V$ (as calculated from Gaia Release 2 distances) versus cos($i$) for our Ritter \& Kolb NL sample whose system parameters are listed in Table 1. Black filled circles are systems having published inclination and uncertainty values which we could verify; empty circles are for {\bf the systems whose inclination and error information come from Ritter and Kolb catalog quoting the unpublished Schafter 1983 PhD thesis; horizontal bars are for those objects with just an inclination constrain in the literature and not a precise determination. The red color denotes} the two old novae RR Pic and OY Ara. 
Dotted and dashed lines represent the expected absolute magnitude for mass transfer rates of 10$^{-8}$ and 10$^{-9}$ M$_\odot$/yr, respectively, as per Webbink et al.'s formula given in the text. The number on the side of each theoretical line indicates the adopted WD mass in solar units. The solid green line is the Warner (1986) relation derived using the Paczynski and Schwarzenberg-Czerny (1980) formulation.}
\end{figure}
 
Using the information in Table 1, we plot the actual M$_V$ values vs. the cosine of the system inclination (Figure 1). {\bf Note: RR Pic and OY Ara are known historical novae; Nova Pic 1925 and Nova Ara 1910, respectively. While it might be more correct to discard these two objects altogether from our sample, we preferred to keep them in since $\simeq$100 yr old nova might be old enough to be considered a quiescent cataclysmic variable (NL). Additionally, recent observations by Sahman et al. (2018) have suggested that V1315 Aql might also have been a nova in the past. In Figure 1, we did not plot CM Del since its inclination uncertainty is so large }

We note a cosine-like dependence on the system brightness vs. $i$.
This is not surprising, and in fact was expected based on our assumption that the luminosity of the disk dominates the light in the optical part of the spectrum. We also see that for a given cos $i$ value, there is a spread of about 2.5 magnitudes in M$_V$.  

Two of the best known and well used M$_V$-Inclination relationships, progenitors of present day relations and usage, are those of Warner (1986) and Webbink et al. (1987).
Warner made use of new absolute magnitude determinations of novae at maximum based on the shell expansion parallax method. He then determined the inclination of each nova remnant by measuring and fitting emission line equivalent widths which he then used as a proxy for viewing angle. Combining these two measured values, available for 13 systems, Warner constructed a plot of M$_V$ vs. cos $i$. While a straight line would have fit his data fairly well, it was desired to see if these data matched the expected M$_V$ vs. cos $i$ relation predicted by theory. Paczynski and Schwarzenberg-Czerny (1980) had developed such a relation based on a disk model they produced for the dwarf nova U Gem during outburst.
Warner used their equation (16), adopting the recommended limb darkening value of $\mu$=0.6, to attempt a match for the distribution of 13 nova remnants in the M$_V$ vs. cos $i$ plane. The Paczynski and Schwarzenberg-Czerny relationship provided only a delta magnitude variation for the geometric aspect effect, 

$$ \Delta M_V (i) = -2.5~log [(1+3/2~cos~i)~cos~i] $$

\noindent making the relationship fully geometric in nature. This expression provides the expected change in M$_V$ when the disk is tilted above or below an inclination of 56.7 degrees ($\Delta M_V$=0), setting $\Delta M_V$ = 4.4 for inclinations $\ge$89 degrees. We present (green line), in Figure 1, Warner's version of the Schwarzenberg-Czerny relation. Warner's nova remnants spanned $\sim$10-100 years after outburst, so the accretion disk in the systems represent a heterogeneous mixture of M$_V$ values. Thus, while this fit follows the general trend in Figure 1, its tie to physical disk parameters is uncertain.

Soon after Warner's publication, Webbink et al. (1987) made a detailed study of recurrent novae and provided in independent view of the relationship between the observed M$_V$ and system inclination. Starting with the basic equations for the luminosity and effective temperature distribution of the standard model accretion disk (c.f., Shakura and Sunyaev, 1973), Webbink et al., convolved the disk luminosity output with a standard $V$ filter and included a geometric inclination dependence to yield 

$$ M_V (obs) = -9.48 -5/3~log(M_{WD} \dot{M}) -5/2~log~(2~cos~i) $$

\noindent where M$_{WD}$ = the mass of the white dwarf in solar masses and $\dot{M}$ is the mass accretion rate of the system in M$\odot$/yr. This relationship provides the expected observed
M$_V$ value based on the physical parameters of the system.

Unlike Warner's empirical relationship, the Webbink et al. equation contains a dependence of the disk luminosity on white dwarf mass and mass accretion rate. Figure 1 shows the Webbink et al. results for the cases of mass accretion rates of 10$^{-8}$ (dotted line) and 10$^{-9}$ M$_\odot$/yr (dashed line), the limiting values expected for these stars, with white dwarf masses of 0.6, 0.8, 1.2 M$\odot$.

The two M$_V$-$cos~i$ relations discussed above provide fairly good fits to the data, revealing the general trend and suggesting that the mass accretion rate is the dominant cause of the spread at any given inclination. But do literature findings agree?

Let us examine a few examples, paying particular attention to the mass accretion rate and the white dwarf mass. 
From the literature we find that the stars MV Lyr (Godon et al., 2017), RW Sex (Hernandez et al., 2017), VY Scl (Hamilton and Sion, 2008), RR Pic\footnote{Note that RR Pic, while listed in the Kolb \& Ritter CV Catalogue as a NL, is an old nova. Such stars have been shown to maintain increased mass accretion rates long after their eruption, possibly the reason for its higher absolute brightness.} (Sion et al., 2017), and DW UMa (Smak 2017) all are shown to have high mass accretion rates, near 10$^{-8}$ M$_\odot$/yr,
while the stars UU Aqr (Dobrotka et al., 2011) and V380 Oph (Zellen et al., 2009) show rates near 
10$^{-9}$ M$_\odot$/yr, the lowest expected for these types of stars. IX Vel (Linnell et al., 2007) and UX UMa (Linnell et al., 2008) have modeled mass accretion rates intermediate to these two groups. 

We note that a change in disk radius (r) between equivalent systems could also cause a change in disk luminosity (L$\propto$r$^{2}$). Using the standard relation between the outer disk radius and $q$ (Lubow \& Shu 1975), we note that even a change in mass ratio from 0.35 (e.g., V425 Cas) to 0.74 (e.g., RW Sex) makes at most a change in r of $\sim$1.4, providing $<$0.5 magnitude of luminosity change for a uniformly bright disk.
Real disks are brighter at smaller r values so, in reality, any minimal areal reduction produces only a small change in disk brightness. 
Additionally, we see no discernible effects in Figure 1 based on mass ratio or orbital period in terms of M$_V$ value. 

We also note that RR Pic, RW Sex, and DW UMa contain high mass white dwarfs, 0.85-0.95 M$_\odot$, placing them at the brightest level for their inclination. Thus, it appears that WD mass does produce a second order, possibly discernible effect. The Webbink et al. model fits shown in Figure 1 seem consistent with the literature values for $\dot{M}$ and M$_{WD}$. 

\section{Summary}
We have used a sample of nova-like systems with orbital periods of 3-6 hr and containing accretion disks that dominate the light in the visible part of the spectrum. Taking the new Gaia DR2 parallax results, we determined the absolute magnitudes for these stars, and with system inclinations taken from the Kolb \& Ritter CV catalog and current references, produced the relationship between M$_V$ and $i$.

Two previous M$_V$-$i$ relationships, based on disk models, initially applied to nova remnants, and still is use today, were examined and proved to be fairly representative of the data.

Webbink et al., (1987) derived a relationship between these two variables that provides a good fit to the observations and allows differentiation between $\dot{M}$ values for a given $i$. We show that the spread in M$_V$, for a given inclination, is indeed dominated by the mass transfer rate (i.e., the disk luminosity being proportional to T$^{4}$; see eq. 5.20, Frank, King, \& Raine 1992) with a small, but perhaps measurable dependence on the white dwarf mass. 
 
Additionally, we confirm that modern theoretical population studies of cataclysmic variables as well as model fits to observational data for individual systems, as discussed in various literature articles, yield derived mass transfer rates consistent with the true M$_V$ for the nova-like stars.


\acknowledgments
We wish to thank Prof. Pierluigi Selvelli for insightful discussions and the anonymous referee for their review which led to a better presentation. EM also thanks Prof. Steven Shore for the confrontation and support. This work has made use of data from the European Space Agency (ESA) mission {\it Gaia} (\url{https://www.cosmos.esa.int/gaia}), processed by the {\it Gaia}
Data Processing and Analysis Consortium (DPAC,
\url{https://www.cosmos.esa.int/web/gaia/dpac/consortium}). Funding for the DPAC has been provided by national institutions, in particular the institutions participating in the {\it Gaia} Multilateral Agreement.


\begin{thebibliography}{}

\bibitem[Corrales(2015)]{2015ApJ...805..23C} Araujo-Betancor, S. et al. 2003, ApJ, 583, 437 
\bibitem[Corrales(2015)]{2015ApJ...805..23C} Bailer-Jones, J.; Rybizki, M.; Fouesneau, G.; et al., 2018, arXiv 1804.10121
\bibitem[Corrales(2015)]{2015ApJ...805..23C} Baptista, R.; Steiner, J. E.; Cieslinski, D., 1995a, ApJ, 433, 332
\bibitem[Corrales(2015)]{2015ApJ...805..23C} Baptista, R.; Horne, K.; Hilditch, R. W.; et al., 1995b, ApJ, 448, 395
\bibitem[Corrales(2015)]{2015ApJ...805..23C} Baptista, R. et al. 1996, MNRAS, 282, 99 
\bibitem[Corrales(2015)]{2015ApJ...805..23C} Bisol, A.; Godon, P.; Sion, E., 2012, PASP, 124 ,158
\bibitem[Corrales(2015)]{2015ApJ...805..23C} Clementini, G., Garofalo, A., Muranena, T., \& Ripepi, V., \ 2018 arXiv 1804.09575
\bibitem[Corrales(2015)]{2015ApJ...805..23C} Dhillon, V. S. et al. 1991, MNRAS, 252, 342 
\bibitem[Corrales(2015)]{2015ApJ...805..23C}Dobrotka, A., Mineshige, S., \& Casares, J. \ 2012, MNRAS, 420, 2467
\bibitem[Corrales(2015)]{2015ApJ...805..23C}
Frank, J., King, A., Raine, D., \ 1992 {\it Accretion Power in Astrophysics}, CAmbridge University Press
\bibitem[Corrales(2015)] {2015ApJ...805..23C}
Gaia Collaboration et al. \ 2018,  2018arXiv180409365G
\bibitem[Corrales(2015)]{2015ApJ...805..23C}Garnavich, P. M. et al. 1990, ApJ, 365, 696 
\bibitem[Corrales(2015)]{2015ApJ...805..23C}Godon, P., et al. \ 2017, ApJ, 846, 52
\bibitem[Corrales(2015)]{2015ApJ...805..23C}Hamilton, R. and Sion E. \ 2008, PASP, 120, 864
\bibitem[Corrales(2015)]{2015ApJ...805..23C}Hernandez, D. \ 2017, MNRAS, 470, 1960
\bibitem[Corrales(2015)]{2015ApJ...805..23C} Howell, S. B., Rappaport, S., \& Nelson, L., \ 2001, ApJ, 550, 897
\bibitem[Corrales(2015)]{2015ApJ...805..23C} Hutchings, J. B et al. 1983, PASP, 95, 264 
\bibitem[Corrales(2015)]{2015ApJ...805..23C} Kolomeni, B., Nelson, L., Rappaport, S., et al., \ 2016, ApJ, 833, 1
\bibitem[Corrales(2015)]{2015ApJ...805..23C} Kolb, U. and Baraffe, I. 2000,NewAR, 44, 99
\bibitem[Corrales(2015)]{2015ApJ...805..23C}Khruzina, T. et al, 2013, A\&A, 511, 125 
\bibitem[Corrales(2015)]{2015ApJ...805..23C}Linnell, A. P., et al. 2005, ApJ, 624, 923 
\bibitem[Corrales(2015)]{2015ApJ...805..23C}Linnell, A., et al. \ 2007a, ApJ, 662, 1204 
\bibitem[Corrales(2015)]{2015ApJ...805..23C}Linnell, A.P., et al. 2007b, ApJ 654, 1036 
\bibitem[Corrales(2015)]{2015ApJ...805..23C}Linnell, A., et al. \ 2008, ApJ, 688, 568
\bibitem[Corrales(2015)]{2015ApJ...805..23C}Lubow, S. H. and Shu, F. H., \ 1975, ApJ, 198, 383
\bibitem[Corrales(2015)]{2015ApJ...805..23C}Luri et al. \ 2018, 20182018arXiv180409376L
\bibitem[Corrales(2015)]{2015ApJ...805..23C}Magnuson, J. A., PhD DARTMOUTH COLLEGE, 1984 
\bibitem[Corrales(2015)]{2015ApJ...805..23C}Mayo, S. K., Wickramasinge, D. T., and Whelan, J. A. J. \ 1980, MNRAS, 193, 793
\bibitem[Corrales(2015)]{2015ApJ...805..23C}Odonoghue, D. et al. 1987, MNRAS, 225, 43 
\bibitem[Corrales(2015)]{2015ApJ...805..23C}Paczynski, B., and Schwarzenberg-Czerny, A. \ 1980, Acta Ast., 30, 127
\bibitem[Corrales(2015)]{2015ApJ...805..23C} Patterson, J. \ 2011, MNRAS, 411, 2695
\bibitem[Corrales(2015)]{2015ApJ...805..23C}Ramsay, G., et al. \ 2017, A\&A, 604, 107
\bibitem[Corrales(2015)]{2015ApJ...805..23C} Ribeiro F. M. A. et al, 2006, PASP, 118, 84 
\bibitem[Corrales(2015)]{2015ApJ...805..23C} Rodríguez-Gil, P., et al. 2000, A\&A, 355, 181 
\bibitem[Corrales(2015)]{2015ApJ...805..23C} Rodriguez-Gil, P. et al. 2004, A\&A 424, 647 
\bibitem[Corrales(2015)]{2015ApJ...805..23C} Rodriguez-Gil, P., et al. 2015, MNRAS, 452, 146 
\bibitem[Corrales(2015)]{2015ApJ...805..23C} 
Sahman, V. S., Dhillon, V. S., Littlefair, S. P., 
\& Hallinan, G., \ 2018, MNRAS, 477, 4483
\bibitem[Corrales(2015)]{2015ApJ...805..23C} 
Schmidtobreick, L. et al. 2018, A\&A in press, arXiv:1806.00097 
\bibitem[Corrales(2015)]{2015ApJ...805..23C} Shakura, N. I. and Sunyaev, R. A. \ 1973, A\&A, 24, 337
\bibitem[Corrales(2015)]{2015ApJ...805..23C} Sion, E., Godon, P., \& Jones, L. \ 2017, AJ, 153, 109
\bibitem[Corrales(2015)]{2015ApJ...805..23C} 
Shafter, A.W. 1983, Ph.D. thesis, UCLA  
\bibitem[Corrales(2015)]{2015ApJ...805..23C} Smak, J., 1994, AcA, 44, 59 
\bibitem[Corrales(2015)]{2015ApJ...805..23C} Smak, J., 1995, AcA, 45, 259 
\bibitem[Corrales(2015)]{2015ApJ...805..23C} Smak, J., \ 2017 AcA, 67, 273
\bibitem[Corrales(2015)]{2015ApJ...805..23C} Thoroughgood, T. D. et al. 2005, MNRAS, 357, 881 
\bibitem[Corrales(2015)]{2015ApJ...805..23C} Vande Putte D. et al. 2003, MNRAS, 341, 151 
\bibitem[Corrales(2015)]{2015ApJ...805..23C}Warner, B. \ 1986, MNRAS, 222,11
\bibitem[Corrales(2015)]{2015ApJ...805..23C}Webbink, R. F., et al. \ 1987, ApJ, 314, 653
\bibitem[Corrales(2015)]{2015ApJ...805..23C} Zhao, P. et al. 1997, ApJ, 483, 899 
\bibitem[Corrales(2015)]{2015ApJ...805..23C}Zellem, R., et al. \ 2009, PASP, 121, 942

\end{thebibliography}
\end{document}